 \documentclass[final,3p,times,twocolumn]{elsarticle}
\journal{Physics Letters A}

\begin{document}

\begin{frontmatter}

 \title{Comments concerning the Ising model and two letters by N.H.~March}

 \author[rvt]{Michael E.~Fisher}

 \author[focal]{Jacques H.H.~Perk\corref{cor1}}
 \ead{perk@okstate.edu}

 \cortext[cor1]{Corresponding author}
 \address[rvt]{Institute for Physical Science and Technology, University of Maryland,\\
    College Park, Maryland 20742-8510, USA}
\address[focal]{Department of Physics, Oklahoma State University,\\
     145 Physical Sciences, Stillwater, Oklahoma 74078-3072, USA}

 \begin{abstract}
  Two recent articles by Norman H.~March that contain misleading statements
  concerning 3D Ising models, partly based on earlier erroneous work of Z.D.~Zhang,
  are addressed.
 \end{abstract}
 

\end{frontmatter}

About a decade ago Zhidong Zhang claimed to have solved the
three-dimensional Ising model exactly and he circulated draft versions. After
several theorists including the two of us had patiently explained errors in the drafts,
Zhang shockingly got an enlarged version published \cite{Zhang}. That the result was in
error was pointed out by Wu, McCoy, Fisher and Chayes \cite{WMFC}, who showed that
Zhang's formula for the free energy could not reproduce the known high-temperature
series expansion while, at low temperatures, his results likewise failed,
disagreeing with C.N.~Yang's exact result for the 2D spontaneous
magnetization, etc. Perk exposed further errors, especially the
erroneous application of the Jordan--Wigner transform \cite{Perk1}. This
should have been the end of it.

However, Zhang next got Norman H.\ March to join him and by now there are more
than two dozen wrong or misleading papers published by them. A detailed invited
commentary on several of these has been published \cite{Perk2} in a Polish
mathematics journal. Zhang then published a review paper in Chinese Physics B,
that was intimidating to some younger scholars in China, as their Monte Carlo
results disagreed with Zhang's work. To help resolve the issue F.Y.~Wu and
one of us were invited to a speak about it at a conference in Beijing, resulting
in another detailed and this time very blunt comment \cite{Perk3}.

However, we were saddened when only a few weeks ago some more recent
papers \cite{March1,March2,March3,Moz} came to our attention. March
still advocates a theory of the critical exponents based on $\alpha=0$
and $\gamma=5/4$ for the three-dimensional Ising model. Even though
these values had been suggested in the long ago past \cite{DS,AF}, they
were based on relatively short series expansions and are no longer
supported by the best research. March thus had to appeal to the
erroneous works of Zhang \cite{Zhang} to support these values.

There is an excellent review by Pelissetto and Vicari \cite{PeVi} that one may consult
for recent expert opinions. The many theoretical and experimental results quoted
in section 3.2 there make it clear that the claims of March are untenable. The results
constitute, however, strong support for the new theory of El-Showk et al.\ \cite{EPPSV}.
This is  an exciting new development in the three-dimensional Ising
model using convex optimization of the $c$-parameter within the conformal
boot-strap approach to the four-point correlation functions. It gives accurate
bounds on the critical exponents that agree with the accepted estimates in the
literature. More precisely \cite{EPPSV,S-D} yield the two scaling dimensions,
\begin{eqnarray}
&&\Delta_{\sigma}   = 0.518151(6) = (1+\eta)/2,\cr
&&\Delta_{\epsilon} = 1.41264(6)  = 3 - 1/\nu,
\end{eqnarray}
from which we calculated
\begin{eqnarray}
&&\nu    = 0.62998(3),\cr
&&\eta   = 0.036302(12),\cr
&&\alpha = 0.11007(7),\cr
&&\beta  = 0.326423(18),\cr
&&\gamma = 1.23708(5),\cr
&&\delta = 4.78982(7),\cr
&&\Delta_{\rm gap} = 1.56351(7) = \beta + \gamma.
\end{eqnarray}
In (1) we have given  the improved bounds obtained by
Simmons-Duffin \cite{S-D}.
The first-correction-to-scaling exponent is \cite{EPPSV}
\begin{eqnarray}
&&\omega = 0.8303(18) = \Delta_{\epsilon'} - 3,\cr
&&\Delta = 0.5231(12) = \omega \nu.
\end{eqnarray}
Note that this value of $\Delta$ is close to $\frac12$, which has been
used for some time in unbiased fits like
\begin{equation}
\chi\sim|t|^{-\gamma}(1+{\rm const}|t|^{\Delta}),\qquad
t=1-{T\over T_{\rm c}},
\end{equation}
for the susceptibility \cite{AF}.

In his articles in \textit{Physics Letters} \cite{March1,March2} March does not
cite the comments on Zhang's work but he refers to them in \cite{March3},
where he gives the impression that someone only has to prove two conjectures
by Zhang, even though these have been disproved already.
March also suggests in \cite{March1,March2} that Zhang's critical exponents
$\alpha=0$ and $\gamma=5/4$ agree within experimental accuracy with all
existing experiments and theory. This is falsified in many works, as discussed above.

Recently March has published two further papers \cite{March4,March5}, the
second with two coauthors, in which he now proposes to accept  $\alpha>0$
but still claims $\gamma=5/4$ to `within both experimental and theoretical ``error"'
and `as known either exactly, or to high accuracy,'
see \cite[p.~14]{March5}. He asserts \cite{March4} that the presently intractable
mathematics of the 3D Ising model was bypassed by Zhang via two conjectures;
however, because of some controversy raised, he now proposes a generalization
with $\alpha\ne0$. Nonetheless we stress that even $\gamma=5/4$ is no longer
compatible with well established and accepted values for the 3D Ising model
such as recorded in (2) above. Besides the review \cite{PeVi} cited above
there are several more recent papers, such as the Monte Carlo study of
Hasenbusch \cite{Has} implying $\gamma=1.23719(26)$
and the experimental study of Sengers and Shanks \cite{SeSh} giving
$\gamma=1.238\pm0.012$. The most accurate experimental values of $\gamma$
are all less than $5/4$ and their average properly weighted with the error bars
is significantly below $5/4$.

In conclusion, the articles by March  in \textit{Physics Letters} \cite{March1,March2}
are doubly misleading: first, because the Zhang exponents are outside the best
experimental and theoretical ranges; and, second, because March does not
mention the comments on Zhang's work that could have alerted the referees.

The authors thank Dr.~Martin Hasenbusch for a very useful comment.

%\section*{References}  %%Delete for arXiv

%\end{linenumbers}

\end{document}